\documentstyle[preprint,pre,aps,epsfig]{revtex}
\begin{document}
\draft
\title{On a new approach to optical solitons in dielectric fibers}
\author{V. Veerakumar\,\,and\,\, M. Daniel$^{\thanks {nld@bdu.ernet.in}}$}
\address{The Abdus Salam International Centre for Theoretical Physics, Trieste, Italy}
\address{\mbox{and}}
\address{Centre for Nonlinear Dynamics, Department of Physics, Bharathidasan 
University, Tiruchirappalli 620 024, India} 
\maketitle 
\begin{abstract}
In this letter we describe a new and simple approach for modulating the 
electromagnetic wave propagating  in a dielectric medium in the form
of solitons by considering the torque developed between the induced 
dipoles in the medium and external field without taking into 
account the nonlinear Kerr effect. A reductive perturbation method
deduces the Maxwell equation coupled with the Larmor equation of the torque
to the derivative nonlinear Schr\"odinger equation that admits
optical solitons in the medium.
\end{abstract}
  
Hasegawa and Tappert \cite{hase}  showed 
that under slowly varying amplitude  
electromagnetic (EM)-pulse propagating in a nonlinear 
Kerr fiber medium is governed by the completely integrable
nonlinear Schr\"odinger (NLS) equation $ iA_x + A_{tt} +\mu|A|^2A=0$ 
that admits N-soliton solutions ($\mu$=constant). This was
derived from Maxwell equations under the assumption of weak linear
dispersion and 
$ A $ is the slowly varying amplitude of the electric field of the 
electromagnetic wave (EMW). 
Later this
 was
experimentally verified by Mollenauer {\it et al}
\cite{moll}.  For more channel handling capacity, it is necessary
to transmit pulses at the order of sub-picosecond and femto-second frequency levels. 
But the propagation of such ultra short pulses (USP) experience 
higher order effects like third order dispersion (TOD), self steepening
(SS) and stimulated Raman scattering (SRS).  
In this case 
the wave propagation is described by the higher order
nonlinear Schr\"odinger (HNLS) equation
$iA_x + A_{tt}$ $+ \mu |A|^2A + i(\mu_1 A_{ttt} $ $ + \mu_2 |A|^2A_t +\mu_3
|A|^2_tA)=0 $. 
For example, when $\mu = 0$ and when the two inertial contributions 
of the nonlinear polarization namely, the 
stimulated Raman scattering (SRS) and self steepening (SS)
are equal ($\mu_2=\mu_3=1$) in the absence of third order dispersion 
(TOD) ($\mu_1=0$), the HNLS
reduces to the completely integrable soliton possessing
derivative nonlinear Schr\"odinger (DNLS) equation \cite{kaup}
$iA_x + A_{tt}$ $+i(|A|^2A)_t=0$, and, however, when 
$\mu \ne 0 $, it reduces to the completely integrable 
mixed derivative nonlinear Schr\"odinger (MDNLS) equation \cite{liu1}
$iA_x + A_{tt}$ $+ \mu |A|^2A$ $+i(|A|^2A)_t=0$ that also admits solitons.
When both the TOD and SS 
effects are included one obtains the Hirota equation
for specific parametric choices ($\mu_1 = 1, \mu_2 = \pm 6, \mu_3=0 $)
\cite{hirota1}
$iA_x + A_{tt}$ $+ \mu |A|^2A + i( A_{ttt}$ $\pm 6|A|^2A_t)=0 $
which explains the propagation of ultra short pulses in the form of soliton. 
Later Sasa and Satsuma \cite{sasa} proved that for a 
particular choice of parametric relations for TOD, SS and SRS 
($\mu_1 = 1, \mu_2 = 6, \mu_3=3$),  the
soliton propagation is supported by another completely integrable HNLS equation
$iA_x + A_{tt}$ $+ \mu |A|^2A + i(A_{ttt} $ $ + 6 |A|^2A_t + 3
|A|^2_tA)=0 $.

The propagation of optical pulses in birefringent fibers have become very
useful in the context of nonlinear directional couplers and a  
lot of work has been carried out recently, in this direction where the dynamical
equations governing the propagation of signals in the form of 
optical solitons reduce to the two coupled nonlinear Schr\"odinger (CNLS) 
family of equations 
\cite{rrk1,rrk2}
$iA_{jx} + A_{jtt}$ $+\mu[ \sum_{k=1}^2 |A_k|^2] A_j $ $+i[ \mu_1 A_{jxxx} + 
\mu_2(\sum_{k=1}^2 |A_k|^2)A_{jx}$ $ +\mu_3 (\sum_{k=1}^2 (|A_k|^2)_x A_j ]=0 $, 
where $j=1,2$.
This has been further extended to $N$-signals \cite{hioe}.
Very recently, the study of  propagation in 
birefringent optical fibers introduced the new concept of shape 
changing solitons that 
share energy amongst themselves during propagation \cite{rrk1,rrk2}. This energy
switching behaviour of optical solitons has been used for constructing
all optical logic gates \cite{jaku}. 
The above models (both single and coupled) support propagation of optical
pulses in the pico- and femto-second ranges that emerge from high intensity
lasers. Now
it has also been experimentally
proved that visible white light  emerging from an
incandescent source, namely a quartz-tungsten-halogen bulb when propagating through 
the photo refractive crystal $Sr_{0.75}Ba_{0.25}Nb_2O_6 $,
optical solitons are formed by self trapping \cite{segev}.
In another direction self modulation of quasi-monochromatic EMW 
into spatially coherent optical solitons
in a dielectric medium 
is described by the
NLS equation  $ iA_t + A_{xx} + 2|A|^2A=0 $ and
their higher order and coupled versions, however with
an interchange of the time($t$) and space($x$) variables in their derivatives 
\cite{zhak,manak}.  

In this letter we describe
a new and simple approach to modulate the EMW propagating 
in a dielectric medium 
in the form of spatially coherent optical solitons 
without taking into account the nonlinear Kerr effect in the 
medium  
as has been done so far but
only by considering the 
torque associated between 
the induced dipole moment of the 
medium and the electric field component of the EM field in the
form of Larmor equation.
When the dielectric medium is exposed to the weak EM field the
induced dipoles, due to polarization when continuously subjected to 
the external field,
give rise to torque. To
the lowest order of perturbation the torque due to the electric field
component of the EM field
is due 
to the interaction of it with the induced electric dipole moment of the 
medium. When the medium is anisotropically polarizable the induced dipole
moment  $ {\bf p} $ and the electric field ${\bf E}$ are not parallel in 
general and the torque is given by ${\bf \Gamma}= {\bf p} \wedge {\bf E} $.
The angular momentum vector is the
only physical quantity that can define a unique direction for the polarised
atom of the dielectric medium.  The classical time evolution equation 
for the angular momentum 
can be constructed by writing the time derivative of the angular momentum 
as equal to the above torque.
Thus we have the gyroscopic equation or the Larmor equation for the 
electric field in the form 
$ {\partial {\bf P}({\bf r},t) \over \partial t} $ 
$= \gamma {\bf P} \wedge {\bf E} $ \cite{hilb}, 
where the polarization ${\bf P} $ is the sum  $ \sum_i{\bf p}_i $ over all the
atoms in a unit volume and $ \gamma $ is the 
Larmor frequency. Even though existence of  anisotropic polarizability 
in the dielectric medium is necessary
for the presence of an electric field induced torque, 
it is not sufficient and therefore in addition, it is required that the 
angular momentum vector should be neither parallel nor orthogonal to the electric field.

The propagation of EMW in the dielectric, when there are no free
stationary and moving electric charges, is described by the Maxwell equations :
\begin{mathletters}
\label{eq2}
\begin{eqnarray}
{\bf \nabla}\cdot{\bf D} = 0, \qquad 
{\bf \nabla}\cdot{\bf B} = 0 \label{eq2a} , \\
{\bf \nabla} \wedge {\bf E} = -{{\partial {\bf B}} \over \partial t}, \qquad
{\bf \nabla} \wedge {\bf H} = {{\partial {\bf D}} \over
\partial t}, \label{eq2b}
\end{eqnarray}
\end{mathletters}
where the fields  ${\bf E} $, ${\bf B} $, ${\bf H} $ and ${\bf D}$, 
which are represented by 3-component vectors, bear the usual text book
meaning and $ \epsilon $ is the dielectric constant of the medium. 
As we consider a dielectric medium 
for EMW propagation here the constitutive relations between 
${\bf B}$ and ${\bf H}$ can be written as 
${\bf H} = 
{{\bf B} \over \mu} $, where $ \mu $ is the magnetic permeability of 
the medium. 
Now taking curl on both sides of the 
first of Eq. (\ref{eq2b}) and using the second of 
Eq.(\ref{eq2b}), and after using the relations for ${\bf H} $ and 
$ {\bf D} = \epsilon
{\bf E} + {\bf P}$, 
we finally obtain
$ {\partial^2 \over {\partial t^2}} 
[{\bf E} + {1 \over \epsilon} {\bf P}] $ $
= c^2 [{\nabla}^2 {\bf E}$ $-{\bf \nabla}
({\bf \nabla}\cdot{\bf E})] $, 
where $ c={ 1 \over {\sqrt{\mu \epsilon }}} $ and ${\bf P} = (P^x,P^y, P^z) $.
To understand the nature of propagation of EMW in the dielectric medium 
we now have to solve Maxwell  equations
coupled with Larmor equation.
As fiber is the convenient form of dielectric medium 
for propagation,
it is more meaningful to study the one dimensional version of these equations 
(say along $x$-direction). We further consider $x$ as the direction of propagation
of EMW, which is taken parallel to the  
length of the dielectric fiber.  Thus we have the following set of coupled
equations to be solved.
\begin{mathletters}
\label{eq4}
\begin{eqnarray}
&&{\partial^2  \over \partial t^2}[ {\bf E}(x,t)  + {1 \over \epsilon} {\bf P}(x,t) ]
= c^2 \left[ {\partial^2 {\bf E}(x,t) \over \partial x^2} 
-  {\partial^2  \over \partial x^2}   E^x (x,t) {\bf e} \right ], \label{eq4a}  \\
&&\hspace{-3cm} \text{and} \nonumber \\
&&{\partial {\bf P}(x,t) \over \partial t} = {\bf P}(x,t) \wedge {\bf E}(x,t), \label{eq4b} 
\end{eqnarray}
\end{mathletters} 
where ${\bf e} = (1,0,0) $ 
and time $t$ has been rescaled and velocity
$ c $ redefined while writing Eq.(\ref{eq4}).

Now, we solve the set of coupled equations (\ref{eq4})
using a reductive perturbation method originally developed by Taniuti and Yajima 
\cite{tanuit} to understand the modulation of the slowly varying EMW due
to nonlinear interactions in the medium.
When the amplitude varies slowly over the period of the 
oscillation, a stretching transformation allows us to separate the system into 
a rapidly varying part associated with the oscillation and a slowly varying one 
such as the amplitude. Nonlinear terms modulate the slowly varying amplitude of the 
wave.  Then a formal solution is given in an asymptotic 
expansion about a uniform value. A similar study has been performed in the context 
of EMW propagation in ferromagnetic medium 
and different types of EM solitons were
obtained depending on the nature of the medium \cite{leblond,kraenk}.
We 
assume that for the set of coupled Eqs.(\ref{eq4}) there exists a formal solution
expanded asymptotically in terms of a small parameter $ \varepsilon $ in the
neighbourhood of a constant value.  For this we expand the components 
of the electric field and polarization of the
medium in terms of a small parameter $\varepsilon $ as
$ {\mathcal{F}}^x (\xi, \tau) $ $= {\mathcal{F}}_0 $ 
$+ {\varepsilon} {\mathcal{F}}^x_1 $ 
$+ {\varepsilon}^2 {\mathcal{F}}^x_2 +$  $ ...,$ 
$ {\mathcal{F}}^{\alpha}(\xi,\tau) $ $= {\varepsilon}^{1 \over 2} $
$[ {\mathcal{F}}^{\alpha}_1 $ 
$+ {{\varepsilon} \mathcal{F}}^{\alpha}_2 $
$+ ...], $
where $\alpha $ $= y, z$. Here
$ {\mathcal{F}} $ stands for the electric field ${\bf E}(x,t)$ as well as
for the polarization
vector $ {\bf P}(x,t) $. 
Due to the anisotropic character of the molecules in the
medium as originally assumed here we have expanded the electric field and 
the polarization
vectors in a nonuniform way about the  constant values $ E_0 $ and $ P_0 $ respectively. 
The expanded electric field and polarization are functions of the slow
variables $ \tau $ and  $ \xi  $ introduced through the stretching  
of time
($ \tau = {\varepsilon}^2t $) and the space variable 
($ \xi = \varepsilon(x - vt)  $), to take
care of the slow variation in amplitude \cite{veera3}.  
Here $\varepsilon $ is the same perturbation parameter introduced earlier
and $ v $ is the
group velocity of the propagating EMW. 

We now substitute the expansions of $ {\bf E} $ and $ {\bf P} $  
in  the component forms of 
Eqs.(\ref{eq4}) 
and collect terms proportional to different powers of $ \varepsilon $
and try to solve the resultant equations. 
On solving the resultant equations 
at $ O( {\varepsilon}^0) $, 
we obtain 
$ E_0 ={-P_0 \over \epsilon} $ and $  
E^{\alpha}_1 = k P^{\alpha}_1 $, where 
$ k \equiv (E_0 / P_0) = 
{c^2 \over {\epsilon(c^2-v^2)}}$. 
Then solving them 
at  $ O( {\varepsilon}^1) $, after using the results of
$ O( {\varepsilon}^0) $ we finally obtain  
$ E^x_1 = {-P^x_1 \over \epsilon} $ and
\begin{mathletters}
\label{eq10}
\begin{eqnarray}
{\partial E^x_1 \over \partial \xi}
&=& {{\hat{\gamma}} \over \epsilon v k} \left[
E^y_1 \int_{-\infty}^{\xi}
{\partial E^z_1 \over \partial \tau} d{\xi'}
- E^z_1 \int_{-\infty}^{\xi}
{\partial E^y_1 \over \partial \tau} d{\xi'} \right], \label{eq10a} \\
{\partial E^y_1 \over \partial \xi}
&=& {{\hat{\gamma}} k P_0 \over v}\int_{-\infty}^{\xi}
{\partial E^z_1 \over \partial \tau}d{\xi'} - {{(1 + \epsilon k)} \over v}
E^z_1 E^x_1, \label{eq10b} \\ 
{\partial E^z_1 \over \partial \xi}
&=&{{(1 + \epsilon k)} \over v}
E^y_1 E^x_1 -  {{\hat{\gamma}} k P_0 \over v}\int_{-\infty}^{\xi}
{\partial E^y_1 \over \partial \tau} d{\xi'}, \label{eq10c} 
\end{eqnarray}
\end{mathletters}
where $ {\hat{\gamma}} = {{2v(2c^2-v^2)} \over {c^2(v^2-c^2)}} $.
Now, we define
\begin{equation}
\label{eq11}
\psi = (E^y_1 - i E^z_1),~~~~~{\mid \psi \mid}^2 =  E^x_1. 
\end{equation}
The above definitions suggest that $ E^x_1 = {E^y_1}^2 + {E^z_1}^2 $, 
which is in accordance with the nonuniform expansion.
Eqs.(\ref{eq10b}) and (\ref{eq10c}), after a single differentiation,
can be  combined together to give after some algebra and rescaling the variable 
${\tau}$ the following 
single equation in terms of the newly defined variable $ \psi $,
\begin{equation}
\label{eq12}
i{\psi}_{\tau} + {\psi}_{\xi \xi} 
+ i \lambda[ {\mid \psi \mid}^2 \psi ]_\xi  = 0.
\end{equation}
where  
$ \lambda = { {-(1 + \epsilon k) k P_0} \over v } $.  
It may be verified that on using the definitions (\ref{eq11}), 
Eq.(\ref{eq10a}) can  be written in the form of Eq.(\ref{eq12}).
Eq.(\ref{eq12}) is the completely integrable derivative
nonlinear Schr\"odinger (DNLS) equation which admits N-soliton solutions.
The multi-soliton solutions of Eq.(\ref{eq12}) can be obtained in the
framework of the inverse scattering transform (IST) method  \cite{kaup}.
It is moreover easy and straightforward to construct the N-solitons
using Hirota's bilinearisation procedure
\cite{hirota}. 
Following ref. \cite{liu}, we consider the nonlinear transformation 
$ \psi = {gf^* \over f^2} $, so that Eq.(\ref{eq12}) can be
written in terms of the following bilinear equations.
\begin{mathletters}
\label{eq13}
\begin{eqnarray}
&& (iD_{\tau} +  D_{\xi}^2 ) g.f = 0, \label{eq13a} \\
&& \left[ (g.f) (iD_{\tau} +  D_{\xi}^2 ) 
+ 2 D_{\xi}(g.f)D_{\xi} \right ] (f^*.f) \nonumber \\
&& - 2 (g.f^*)D_{\xi}^2(f.f) + i \lambda \left[ 3 D_{\xi}(g.f) + gf D_{\xi}
\right] (g^*.g) = 0, \label{eq13b}
\end{eqnarray}
\end{mathletters}
where $ D_{\tau}^m$ $ D_{\xi}^n $ $(g \cdot f) $ $= \left[ 
{\partial \over \partial \tau} - {\partial \over \partial \tau'} \right]^m  $ $\left[ 
{\partial \over \partial \xi} - {\partial \over \partial \xi'}\right]^n $ 
$ g(\xi,\tau) \cdot f(\xi',\tau')|_{\mbox{at}~\xi'=\xi,~\tau'=\tau} $ are Hirota's 
bilinear operators \cite{hirota}. 
In order to obtain soliton solutions, we assume
the series expansions for $ g $, $ f $  and $ f^*$ as $ g=\chi g_1 + \chi^3 g_3 +...$, 
$f= 1 + \chi^2 f_2 + \chi^4 f_4 + ... $ and 
$f^*= 1 + \chi^2 f^*_2 + \chi^4 f^*_4 + ... $ where $ \chi $ is an arbitrary
parameter.  Now, for example to construct a one-soliton solution we set 
$ g = \chi g_1 $, $ f= 1+\chi^2 f_2  $ and $ f^*= 1+\chi^2 f^*_2  $
and then collect terms with similar powers of $ \chi $ and solve the
resultant equations which admit the following solutions for $ g $, $ f $ and
$ f^* $.
\begin{mathletters}
\label{eq36}
\begin{eqnarray}
g(\xi,\tau) &=& exp[\eta_1]. \label{eq36c} \\
f(\xi,\tau) & = & 1 + 
{{i\lambda \Omega_1} \over {2 (\Omega_1 + \Omega_1^*)^2} }
exp[\eta_1 + \eta_1^*], \label{eq36a} \\
f^*(\xi,\tau) & = & 1 -
{{i\lambda \Omega_1^*} \over { 2 (\Omega_1 + \Omega_1^*)^2}}
exp[\eta_1 + \eta_1^*], \label{eq36b} 
\end{eqnarray}
\end{mathletters}
where  $ \eta_{1} \equiv \eta_{1R} + i \eta_{1I} $ $= K_1 \tau + \Omega_1 \xi + \eta_1^{(0)} $, 
$K_1 = i \Omega_1^2 $ and  $ \Omega_1 = \Omega_{1R}
+i \Omega_{1I} $,
and the constant $ \eta_1^{(0)} = \eta_{1R}^{(0)} + i \eta_{1I}^{(0)} $.
Thus the one soliton solution for $ \psi ={gf \over f^*} $ is found to be
\begin{equation}
\psi = Q sech(\eta_{1R} +A_0) tanh( \eta_{1R} +A_0),
\label{sol}
\end{equation}
where $ Q = {-1 \over 2} exp[ i \eta_{1I} + A ] $, 
$ A_0 = \eta_{1R}^{(0)}+A $ and  $ A={1 \over 2} 
ln({{i \lambda \Omega_1} \over 8 \Omega_{1R}^2}) $.
Similarly we can find two, three ... and N-soliton solutions. 
As the details of constructing multi-soliton solutions are very lengthy 
and the form of $N$-soliton solution is very cumbersome
we are
not presenting the details here and interested readers can refer to
ref. \cite{liu}.
Using the one soliton solution given in 
Eq.(\ref{sol}) in the definition of $ \psi $ (Eqs.(\ref{eq11})) the 
components of the electric field vector can be constructed.  Thus the
optical one soliton in terms of the electric field vector component
of the EM field can be written as
\begin{mathletters}
\label{elect}
\begin{eqnarray}
E^x_1 &=&  {-P^x_1 \over \epsilon } = 
{1 \over 4}exp(2A)
sech^2(\eta_{1R} +A_0) tanh^2( \eta_{1R} +A_0), \\
E^{\alpha}_1 &=& k P^{\alpha}_1 = -{1 \over 4} exp(A) [ Re(Q) \delta_{x\alpha}
- iIm(Q)\delta_{y\alpha}]sech(\eta_{1R} +A_0) tanh( \eta_{1R} +A_0),
\end{eqnarray}
\end{mathletters}
where $ Re(Q) $ and $ Im(Q) $ represent the real and imaginary parts of $ Q $ 
respectively and $\delta_{x\alpha}$ and $\delta_{y\alpha}$ are Kronecker delta
functions. 

Eqs.(\ref{elect}) show that the electric field component of the 
propagating plane EMW is modulated in the form of solitons by
compensating the dispersion with the torque produced by the 
rotation of the dipoles induced in the medium. In both Hasegawa's 
theoretical model 
\begin{figure}[h]
\begin{center}
\epsfig{figure=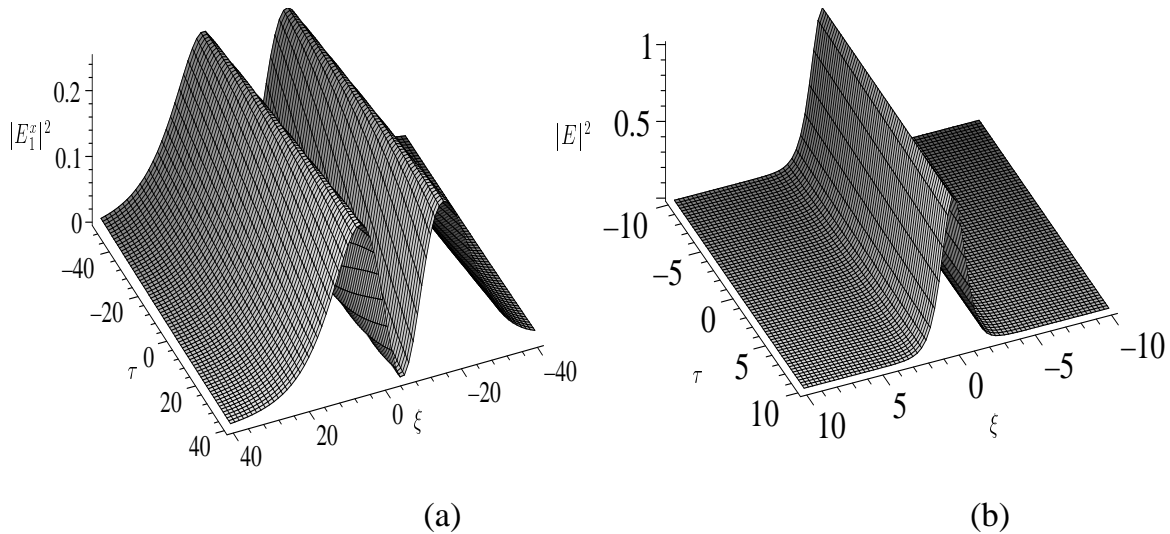, width=\columnwidth}
\end{center}
\caption{(a)
x-component of $ E_1 (E^x_1)$ exhibiting soliton behaviour with two maxima 
(twin pulses)
when
$\Omega_{1R}=0.08$, $ \Omega_{1I}=0.003 $,
$ \eta_{1R}=0.02 $ , $ \lambda = 0.1 $ and $ A = 0.3010 $, 
(b) the form of NLS one-soliton for comparison.} 
\end{figure}
\noindent
for optical soliton in the pico-second region (temporal coherence) and in the case
of Zhakarov/Manakov's model for modulation of the EMW in the
form of solitons (spatial coherence), the soliton is governed by
NLS equation. 

However, in  our problem where the Kerr nonlinearity is not
taken into account, but, by getting support for nonlinearity from the rotation 
of induced dipoles in the medium due to the torque with the external field
the soliton is governed by the DNLS equation. 
The spatial soliton is formed when the EMW induces a waveguide in the
dielectric medium via the nonlinearity of the torque induced, 
self trapped and in turn is guided in its own wave guide. 
Unlike the NLS-one soliton, here we have multihump (two) soliton
(see Figs.(1a) and (1b)). In nonlinear optics, multihump 
solitons corresponding to multimodes propagating without 
interference and loosing their stability are known
\cite{segev1,ostro}.  Even though we get multihump (two humps)
solitons, in our case it corresponds to a single mode which is
again a stable soliton corresponding to the completely integrable
DNLS equation.

Now to understand the effect of the nonlinear term ($|\psi|^2\psi)_{\xi} $) in
the optical soliton formation during propagation of EMW we consider
Eq.(\ref{eq12}) by dropping the dispersion term \cite{ander}.
\begin{equation}
\label{anal}
\psi_{\tau} + \lambda (|\psi|^2\psi)_{\xi} = 0.
\end{equation}
Assuming the solution in the form of $\psi(\xi,\tau)$ $= q(\xi,\tau) 
exp[i\theta(\xi,\tau)]$, Eq.(\ref{anal}) can be written after 
rescaling $\tau $ suitably as 
$ q_{\tau} $ $+ 3 q^2 q_{\xi} = 0 $ ,
~$ \theta_{\tau} $ $+ q^2 \theta_{\xi} = 0 $,
for which the solution can be written in the implicit form.
Infact the first of the above equations is equivalent
to the dispersionless
modified Korteweg - de Vries (mKdV) equation for which the solution can
be written in the implicit form as 
$ q^2=a(\xi - 3\tau q^2) $
or equivalently in another form as $ \xi = 3\tau q^2 + b(q^2) $, where
$ a$ is an arbitrary function determined by the initial profile
and $ b $ is the inverse function $ b = {1 \over a} $.
The second equation of the above equations 
can be solved using the method of characteristics 
and again the solution can be written in terms of another arbitrary
function. 
Explicit solutions can be found for specific initial pulse
profiles.  For example, if we choose a Gaussian pulse in the
form  $ q^2(\xi,0) = exp({-\xi^2 \over \xi_0^2}) $ and
using this, we find that $ a(\xi) = exp({-\xi^2 \over \xi_0^2}) $ 
and 
hence $ b(q^2) = \pm \xi_0 
(ln{1 \over \rho^2})^{1 \over 2} $. This leads to the solution 
$ q^2 = exp[-{(\xi-3\tau q^2)^2 \over \xi^2} ] $ which gives
$ \xi = 3\tau q^2 \pm \xi_0(ln{1 \over \rho^2})^{1 \over 2} $,
where the signs $ (+) $ and $ (-) $ refer to the trailing and 
leading edges of the soliton
respectively. From the above 
expression  for $ \xi $ it is clear that the local
increment in velocity is inversely proportional to the square of the amplitude
thus leading to a 
shift of the pulse peak to the trailing edge. 
As in the case of mKdV equation here also the
self-steepening of the trailing edge leads to the appearance of an 
EM shock wave. If we repeat the calculations for a sech-profile in the
form $q^2(\xi,0) = sech^2({\xi \over \xi_0}) $, 
the results show the formation of a shock EMW at 
the trailing edge of the pulse as in the Gaussian case.  
Further, if we choose a periodic wave profile 
in the form $q^2(\xi,0) = cos({\xi \over \xi_0}) $, once again we
obtain the same result exhibiting shocks in the negative slope region
of the propagating EMW.  This EM shock is compensated by the dispersion
to form a localized spatially coherent optical soliton. 

In this paper we have proposed a simple and new approach to generate
optical solitons in a dielectric medium during propagation of EMW without
taking into account the nonlinear Kerr effect as done normally. On the 
other hand we considered  the torque 
produced in the medium due to the rotation of the induced dipoles
by the interacting exterior electric field component of the EM field
and constructed the Larmor equation for precession of the induced dipoles 
(or electric polarization) in the medium. This is then solved with
Maxwell equations  which govern the propagation of EMW in the dielectric 
medium using a reductive perturbation method. The equations were reduced to 
the integrable derivative nonlinear Schr\"odinger equation that admits 
$N$-soliton solutions. Here the nonlinearity required to support optical soliton 
is provided by the torque developed between the induced electric dipoles in the 
medium and the EM field and not from the Kerr effect.  Another important 
difference, when comparing the results of the usual optical soliton study, 
is that, unlike the other case here we get multihumped optical solitons of the 
DNLS equation.  A careful analysis of the final 
results showed that the nonlinearity due to the torque actually produced a shock in 
the propagating EMW and when it is balanced by the linear dispersion, the plane EMW 
is modulated into spatially coherent optical solitons. Also, the polarization of 
the medium is excited in the form of solitons. We 
expect our method to bring out fascinating results of energy sharing optical 
solitons and optical logic gate operations in birefringent fiber medium in a 
straightforward way through this analytic approach. Further, this is expected
to be useful in analysing all higher order effects in polarization of the medium.

This work was done within the framework of the Associateship Scheme of the 
Abdus Salam International Centre for Theoretical Physics, Trieste, Italy. 
V.V acknowledges  CSIR for financial support in the form of a Senior Research
Fellowship.The work of M.D forms part of a major DST project.

\end{document}